\documentclass[aps,prl,amsmath,amssymb,reprint,twocolumn,superscriptaddress,showpacs]{revtex4}
\usepackage{epsfig,amssymb}
\usepackage{graphicx}
\usepackage{dcolumn}
\usepackage{bm}
\usepackage{multirow}
\begin{document}

\title{Effects of self-consistency and plasmon-pole models on GW calculations for closed-shell molecules}

\author{Johannes~Lischner}
\email{jlischner@civet.berkeley.edu}
\affiliation{Department of Physics, University of California,
  Berkeley, California 94720, USA}
\affiliation{Materials Sciences Division,
  Lawrence Berkeley National Laboratory, Berkeley, California 94720, USA.}
\author{Sahar~Sharifzadeh}
\affiliation{Molecular Foundry, Materials Sciences Division,
  Lawrence Berkeley National Laboratory, Berkeley, California 94720, USA.}
\author{Jack Deslippe}
\affiliation{NERSC, Lawrence Berkeley National Laboratory, Berkeley, California 94720, USA.}
\author{Jeffrey B. Neaton}
\affiliation{Department of Physics, University of California,
  Berkeley, California 94720, USA}
\affiliation{Materials Sciences Division, Lawrence Berkeley National
  Laboratory, Berkeley, California 94720, USA.}
\affiliation{Molecular Foundry, Materials Sciences Division,
  Lawrence Berkeley National Laboratory, Berkeley, California 94720, USA.}
\affiliation{Kavli Energy NanoSciences Institute at Berkeley, Berkeley, California 94720, USA}
\author{Steven G. Louie}
\affiliation{Department of Physics, University of California,
  Berkeley, California 94720, USA}
\affiliation{Materials Sciences Division, Lawrence Berkeley National
  Laboratory, Berkeley, California 94720, USA.}

\begin{abstract}
  We present theoretical calculations of quasiparticle energies in
  closed-shell molecules using the GW method. We compare three
  different approaches: a full-frequency $G_0W_0$ (FF-$G_0W_0$) method
  with density functional theory (DFT-PBE) used as a starting mean
  field; a full-frequency $GW_0$ (FF-$GW_0$) method where the
  interacting Green's function is approximated by replacing the DFT
  energies with self-consistent quasiparticle energies or Hartree-Fock
  energies; and a $G_0W_0$ method with a Hybertsen-Louie generalized
  plasmon-pole model (HL GPP-$G_0W_0$). While the latter two methods
  lead to good agreement with experimental ionization potentials and
  electron affinities for methane, ozone, and beryllium oxide molecules,
  FF-$G_0W_0$ results can differ by more than one electron volt from
  experiment. We trace this failure of the FF-$G_0W_0$ method to the
  occurrence of incorrect self-energy poles describing shake-up
  processes in the vicinity of the quasiparticle energies.
\end{abstract}

\maketitle

\emph{Introduction}.---Accurate knowledge of the energy of
quasiparticle excitations is necessary to interpret photoemission
\cite{LischnerVigil,Rubio}, inverse photoemission,
tunnelling\cite{Dial}, transport \cite{MolTransport} and other
single-particle excitation experiments. The determination of
quasiparticle energies is also an important step in the calculation of
optical absorption and reflectivity spectra \cite{LouieRohlfing}.

The GW method \cite{HedinBook,LouieHybertsen}, in which the electron
self energy is evaluated to first order in the screened Coulomb
interaction $W$ and the one-electron Green's function $G$, is the
current state-of-the-art approach for calculating accurate
quasiparticle energies in crystalline bulk solids, surfaces and
nanostructures from first principles. To simplify such calculations,
additional approximations are often invoked. Most studies employ a
one-shot procedure, where the self energy is evaluated using the
Green's function and screened Coulomb interaction from a DFT
mean-field calculation. In addition, many studies employed generalized
plasmon-pole models \cite{LouieHybertsen,GodbyNeeds,LindenHorsch} to
avoid the explicit calculation of the screened interaction at non-zero
frequencies.

In recent years, many studies have applied the GW method to molecular
systems
\cite{Bechstedt,LischnerOpen,SaharNeaton,SteveGrossmann,NeatonLouie,BlaseOlevano,SanchezPortal,ThygesenRubio,ThygesenOlsen}.
Despite these efforts, it is not yet clear to what degree the
approximations which are commonly used in GW calculations on extended
systems are valid or effective in molecular systems. Previous studies
explored the dependence of the results of one-shot GW calculations on
the mean-field starting point
\cite{Marom,ThygesenRostgaard,ShirleyMartin}. Other studies
investigated the effect of self-consistency by iterating Hedin's
equations, but neglected vertex corrections
\cite{Caruso,ThygesenRostgaard,vanLeeuwen}. Also, several works on
molecules employed a generalized plasmon-pole model
\cite{SaharNeaton,SteveGrossmann}. Plasmon-pole models were originally
introduced for calculations on the homogeneous electron
gas\cite{HedinBook}, where the inverse dielectric function exhibits a
single, sharp plasmon peak, and later extended to crystals using
additional sum rules \cite{LouieHybertsen}.

In this article, we explore the importance of self-consistency and the
validity of generalized plasmon-pole models in GW calculations for
molecular systems. Instead of focusing on quasiparticle energies, we
investigate the frequency-dependent self energies. We observe that the
self energies exhibit many poles whose positions depend sensitively on
the degree of self-consistency used in the GW calculation. These poles
describe shake-up processes, where in addition to the quasiparticle an
electron-hole pair is created \cite{cederbaum}. In non-selfconsistent
calculations with a DFT starting mean field, we find that self-energy
poles can occur erroneously close to the quasiparticle energies
leading to \emph{significant disagreement with experiment} for such
excitations. Including effects of self-consistency by replacing the
DFT-PBE orbital energies by self-consistent quasiparticle energies ---
or equivalently for molecules by Hartree-Fock energies --- moves the
self-energy poles away from the quasiparticle energies and gives good
agreement with experiment. Remarkably, we find that non-selfconsistent
calculations employing a generalized plasmon-pole model
\cite{LouieHybertsen} that conserves sum rules also yield accurate
results.

\emph{Methods}.---The energies $E_n$ of quasiparticle excitations are
the poles of the interacting one-electron Green's function and can be
calculated by solving the quasiparticle or Dyson's equation
\begin{align}
  h(\bm{r})\Psi_n(\bm{r}) + \int d\bm{r'}
  \Sigma(\bm{r},\bm{r'},E_n)\Psi_n(\bm{r'}) = E_n \Psi_n(\bm{r}),
  \label{eq:qp}
\end{align}
where
$h(\bm{r})=-\frac{1}{2}\nabla^2+V_{ion}(\bm{r})+V_H(\bm{r})$. Here,
with $V_{ion}$ and $V_H$ denote the ionic potential and the Hartree
potential, respectively, and $\Psi_n$ is the quasiparticle wave
function. And, $\Sigma$ is the electron self energy, which we
calculate in the $GW$ approximation as
\begin{align}
  \Sigma(\bm{r},\bm{r}',\omega) = i\int
  \frac{d\omega'}{2\pi}e^{-i\eta \omega'}
  G(\bm{r},\bm{r}',\omega-\omega') W(\bm{r},\bm{r}',\omega')
  \label{eq:sigma}
\end{align}
with $\eta=0^+$. As mentioned, $G$ denotes the interacting Green's
function and $W$ the screened Coulomb interaction.

Expressing Eq.~\eqref{eq:qp} in the basis of mean-field
orbitals $\psi_n$ and neglecting off-diagonal matrix elements of the
self energy, the quasiparticle equation becomes
\begin{align}
  E_n = \epsilon_n + \Sigma_n(E_n) -  V^{xc}_n,
  \label{eq:dyson}
\end{align}
where $\epsilon_n$ and $V^{xc}_n$ denote the orbital energies and
exchange-correlation potential matrix elements from a mean-field
theory calculation and $\Sigma_n(E_n)=\langle \psi_n |
\Sigma(E_n)|\psi_n \rangle$. 

In practice, $G$ and $W$, which are needed to construct $\Sigma$, must
be evaluated within certain approximations. In the $G_0W_0$
approximation, one uses $G$ and $W$ from a mean-field calculation.

Going beyond the $G_0W_0$ approximation is challenging. In principle,
one could iterate Eqns.~\eqref{eq:dyson} and \eqref{eq:sigma} and
recalculate $G$ and $W$ using the quasiparticle energies. However,
because of the neglect of the vertex corrections, this procedure is
not guaranteed to converge accurately to the physical result \cite{Holm}. Another
possibility is to update only the Green's function in Eq.~
\eqref{eq:sigma}, while keeping the screened interaction $W_0$ from a DFT
mean-field theory. This method is motivated by the observation that,
for many molecular and other large band gap systems, the mean-field
energies from DFT-PBE differ significantly from the experimental
quasiparticle energies. DFT-PBE \emph{energy differences}, however,
are often serendipitously close to neutral excitation energies (see below),
which are the poles of the screened interaction. This method, the
$GW_0$ approximation, can yield excellent results for both molecular and
extended systems\cite{Holm,vanLeeuwen}.

Even with the $G_0W_0$ approximation, the calculation of the self
energy for molecules is computationally challenging. To evaluate the
frequency integral in Eq.~\eqref{eq:sigma}, it is necessary to compute $G$
and $W$ on a sufficiently fine frequency grid.  Each evaluation of $W$
requires a sum over all empty states to calculate the polarizability
and then a matrix inversion to obtain its inverse. To reduce the
computational effort, a generalized plasmon-pole model is often used
to extend the zero-frequency inverse dielectric matrix to finite
frequencies \cite{LouieHybertsen,SaharNeaton,NeatonLouie}.

The generalized plasmon-pole model of Hybertsen and Louie
\cite{LouieHybertsen} assumes the inverse dielectric matrix
($\omega>0$) can be expressed as
\begin{align}
  \text{Im} \epsilon^{-1}_{\bm{G}\bm{G}'}(\omega) = A_{\bm{GG}'}
  \delta(\omega - \tilde{\omega}_{\bm{GG}'}),
\end{align}
where $\bm{G}$ and $\bm{G}'$ are reciprocal lattice vectors (we assume
a periodic supercell approach) and $\tilde{\omega}_{\bm{GG}'}$ denotes
an effective excitation energy.  Both $A_{\bm{GG}'}$ and
$\tilde{\omega}_{\bm{GG}'}$ are determined by imposing the f-sum rule
and the Kramers-Kronig relation \cite{LouieHybertsen}.

\emph{Computational details}.---We calculate self energies and
quasiparticle properties for the beryllium oxide (BeO) molecule,
methane (CH$_4$), and ozone (O$_3$).  We first carry out DFT
calculations with the PBE exchange-correlation functional, a plane
wave basis, and norm-conserving pseudopotentials. For this, we employ
the QUANTUM ESPRESSO program package \cite{QuantumEspresso}. We then
calculate the quasiparticle energies in the full-frequency $G_0W_0$
(FF-$G_0W_0$) approximation using a basis of Kohn-Sham
orbitals\cite{LischnerOpen,Tiago,ShirleyMartin}. Because of the large
computational expense, carrying out self-consistent FF-$GW_0$
calculations is challenging. To approximate the result of a FF-$GW_0$
calculation, we update the DFT-PBE energies by solving
Eq.~\eqref{eq:dyson} with the Hartree-Fock approximation for the self
energy and use the resulting Green's function, which still has a
simple quasiparticle form, in Eq.~\eqref{eq:sigma}. Because screening
is weak in a molecule, the Hartree-Fock energies are often much closer
to the final quasiparticle values than DFT-PBE energies, and the
Hartree-Fock Green's function is a good approximation to the
self-consistent interacting Green's function.  Finally, we compute the
G$_0$W$_0$ self energy using the generalized plasmon-pole
approximation of Hybertsen and Louie (denoted HL GPP-$G_0W_0$). For
all $GW$ calculations, we employ the BerkeleyGW program package
\cite{BGWpaper}.

To obtain converged results, we use 950 empty states in the
calculation of the screened interaction and the self energy. In
addition, we employ a static remainder correction to approximately
include the effects of missing unoccupied states in the self energy
\cite{StaticRemainder}. In the calculation of the screened
interaction, we use supercell reciprocal lattice vectors of kinetic
energy up to 12 Ry (CH$_4$), 24 Ry (BeO) and 30 Ry (O$_3$).  Finally,
we employ a truncated Coulomb interaction to avoid interactions
between periodic replicas\cite{sohrab}.

\begin{figure}
  \includegraphics[width=8.cm]{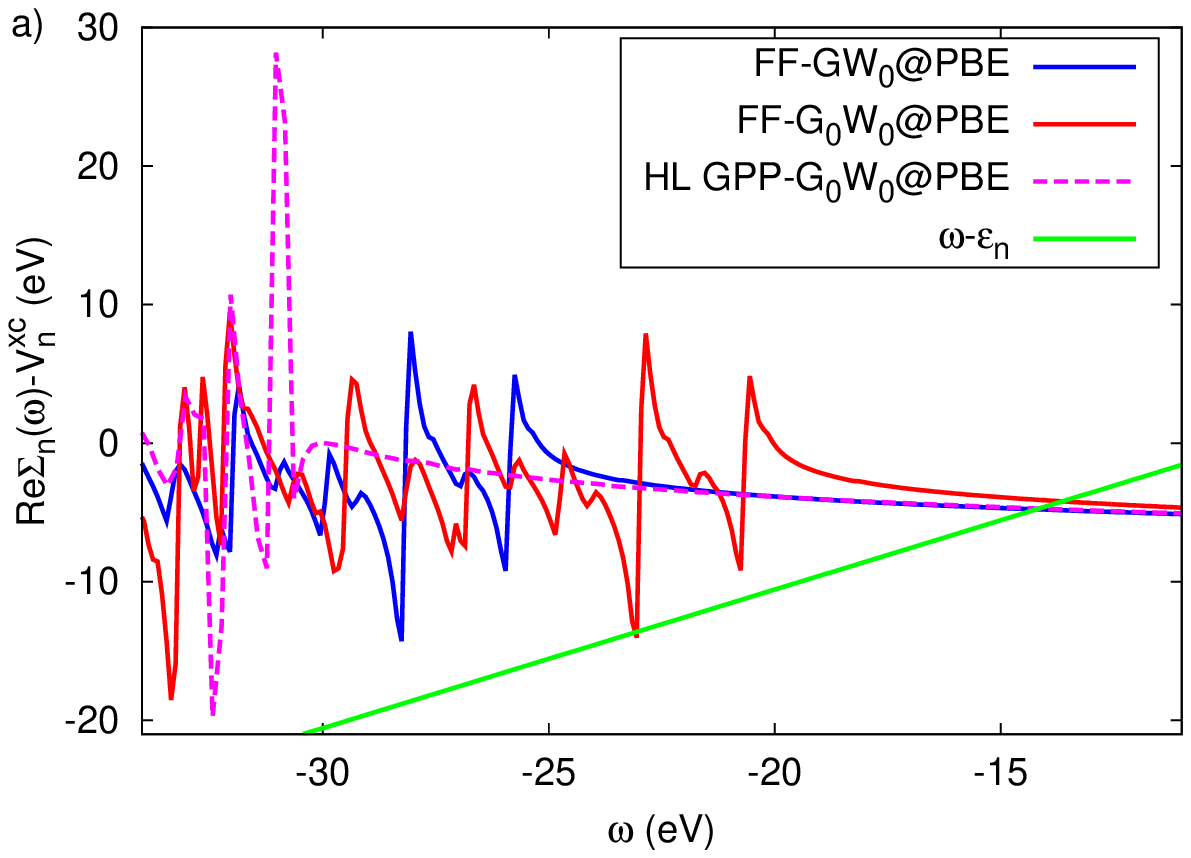} 
  \includegraphics[width=8.cm]{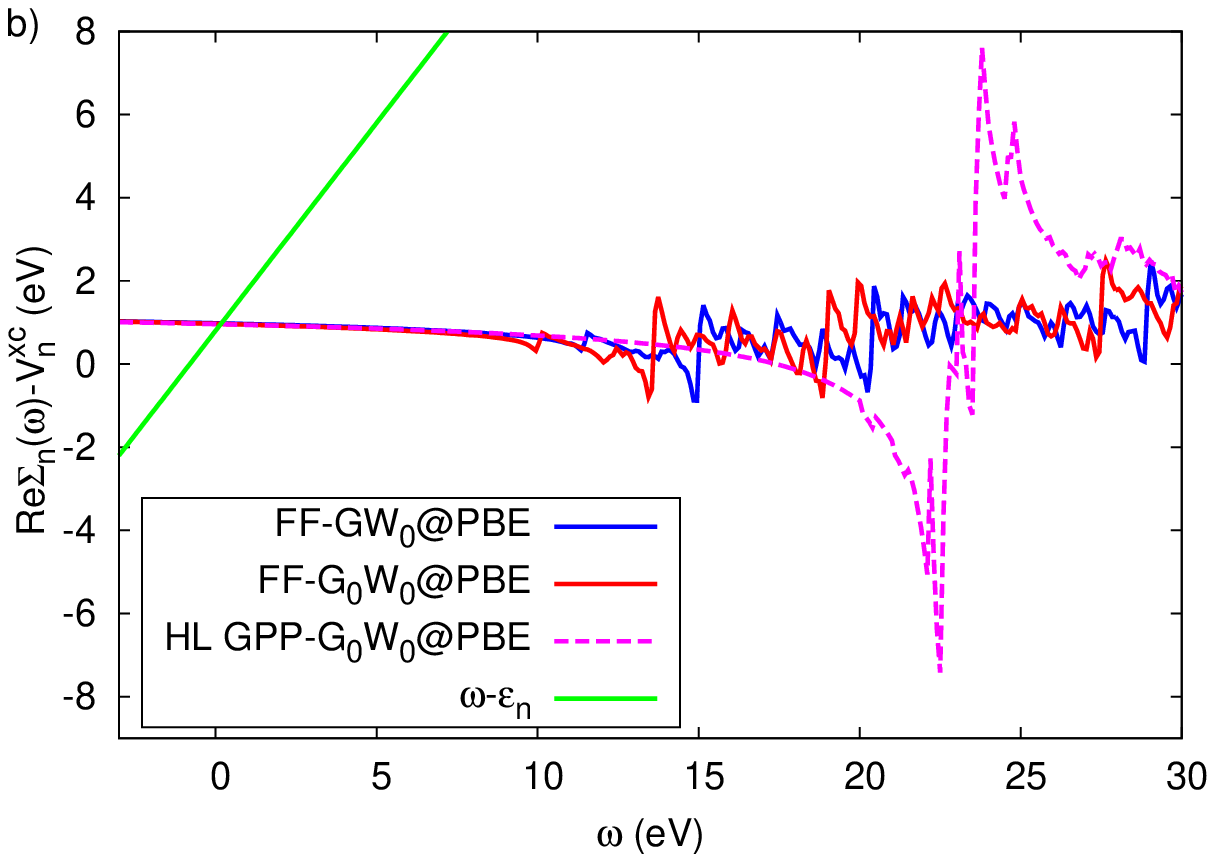}
  \caption{Graphical solution of the quasiparticle equation for the HOMO (a) and
    the LUMO (b) of methane. The quasiparticle energies $E_n$ are
    given by the values of $\omega$ at the intersections of
    $\omega-\epsilon_n$ and
    $\text{Re}\Sigma_n(\omega)-V^{xc}_n$. Shown are self energies from
    full-frequency G$_0$W$_0$ theory, full-frequency GW$_0$ theory and
    G$_0$W$_0$ theory with the generalized Hybertsen-Louie plasmon-pole
    approximation. All calculations employed a DFT-PBE starting point.}
  \label{fig:CH4}
\end{figure}

\emph{Results}.---Figure~\ref{fig:CH4}(a) shows the graphical solution
of the quasiparticle equation for the highest occupied molecular
orbital (HOMO) of the CH$_4$ molecule from the FF-G$_0$W$_0$,
FF-GW$_0$ and HL GPP-G$_0$W$_0$ approaches. All self energies are
smooth functions of frequency in the vicinity of the quasiparticle
solution. At more negative energies, the self energies exhibit many
poles. The onset of these singularities occurs at \emph{less negative}
energies in the FF-G$_0$W$_0$ method with the first pole occurring at
$\sim -21$~eV. The slower decay of the corresponding tail leads to a
$\sim 0.55$~eV difference of the HOMO energy compared to FF-GW$_0$ and
HL GPP-G$_0$W$_0$, which agree very well with each other and with
experiment (see Table~\ref{table:energies}).

Figure~\ref{fig:CH4}(b) shows the self energies associated with the
lowest unoccupied orbital (LUMO) of CH$_4$. Here, no poles of the self
energy are located in the vicinity of the quasiparticle solution and
all three approaches are in good agreement.

\begin{table*}[ht]
  \setlength{\doublerulesep}{1\doublerulesep}
  \setlength{\tabcolsep}{2\tabcolsep}
  \caption{Comparison of quasiparticle energies from various
    theoretical approaches with experiment\cite{CRC}: DFT-PBE, Hartree-Fock
    (HF), full-frequency $G_0W_0$ (FF-$G_0W_0$), full-frequency $GW_0$
    (FF-$GW_0$) and $G_0W_0$ with the Hybertsen-Louie generalized plasmon-pole approximation
    (HL GPP-$G_0W_0$). In all calculations, a DFT-PBE starting mean field was
    employed. All energies are given in eV.}
  \begin{ruledtabular}
    
    \begin{tabular}{c c| d d d d d d}
     &  & \multicolumn{1}{c}{DFT-PBE} & \multicolumn{1}{c}{HF} & \multicolumn{1}{c}{FF-$G_0W_0$@PBE} & \multicolumn{1}{c}{FF-$GW_0$@PBE} & \multicolumn{1}{c}{HL GPP-$G_0W_0$@PBE} &
     \multicolumn{1}{c}{Exp.} \\
      \hline
     CH$_4$ &HOMO &  -9.44  &  -14.63  & -13.64 & -14.21 & -14.16 & -14.35 \\
     CH$_4$ &LUMO  &  -0.80  &      0.60  &     0.16 &     0.18
                  & 0.16 & -\\
      \hline
      BeO &HOMO & -6.24 & -11.35 & -8.76 & -10.46 & -10.56 & -10.1\\
      BeO &LUMO & -4.83 & -0.88 & -2.65 & -2.16 & -2.41 & -\\
       \hline
       O$_3$ &HOMO & -7.96 & -14.31 & -11.43 & -12.97 & -12.72 & -12.73\\
       O$_3$ &LUMO & -6.16 & -1.07 & -2.53 & -2.55 & -1.86 & -2.10\\
    \end{tabular}

  \end{ruledtabular}
  \label{table:energies}
\end{table*}

\begin{figure}
  \includegraphics[width=8.cm]{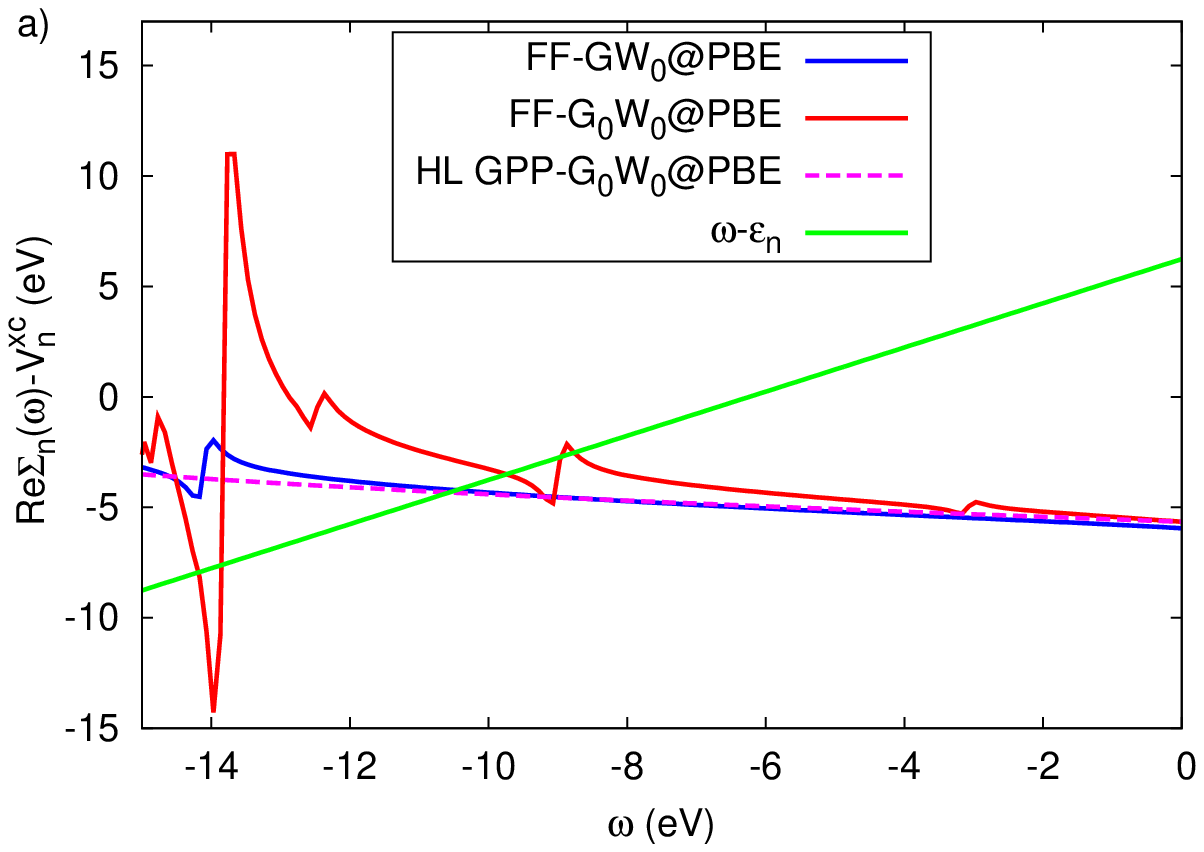} 
  \includegraphics[width=8.cm]{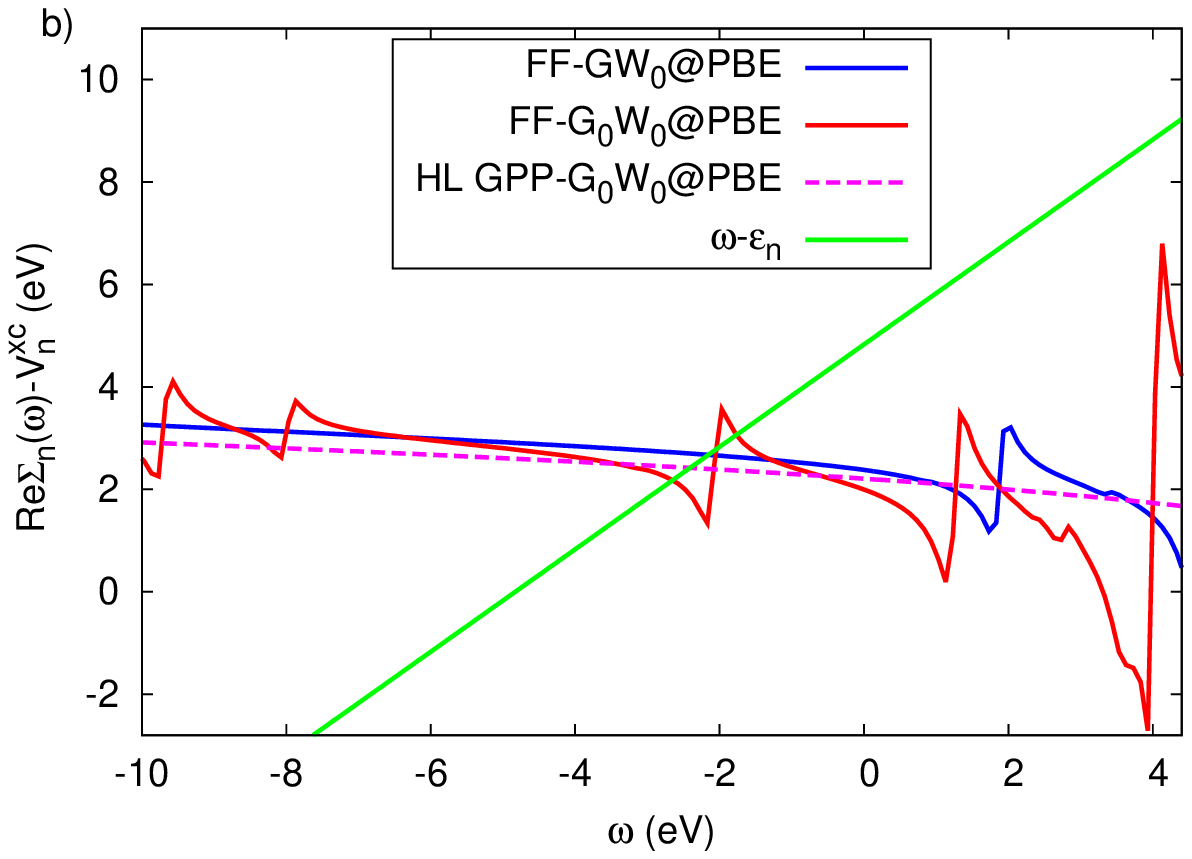}
  \caption{Graphical solution of the quasiparticle equation for the HOMO (a) and
    the LUMO (b) of the beryllium oxide molecule. The quasiparticle energies $E_n$ are
    given by the values of $\omega$ at the intersections of
    $\omega-\epsilon_n$ and
    $\text{Re}\Sigma_n(\omega)-V^{xc}_n$. Shown are self energies from
    full-frequency G$_0$W$_0$ theory, full-frequency GW$_0$ theory and
    G$_0$W$_0$ theory with the generalized Hybertsen-Louie plasmon-pole
    approximation. All calculations employed a DFT-PBE starting point.}
  \label{fig:BeO}
\end{figure}

Figure~\ref{fig:BeO}(a) shows the graphical solution of the
quasiparticle equation for the HOMO of BeO. The FF-$G_0W_0$ solution
nearly coincides with a pole of the self energy, while for the other
methods the self-energy poles are located at more negative energies
and the quasiparticle solution occurs in a region where the self
energy is smooth. The FF-$GW_0$ result differs from experiment by
$0.36$~eV and agrees well with the HL GPP-$G_0W_0$ result. In
contrast, the FF-$G_0W_0$ quasiparticle energy differs from experiment
by $1.34$~eV. A similar situation occurs for the LUMO, see
Fig.~\ref{fig:BeO}(b). Again, the FF-$G_0W_0$ quasiparticle solution
nearly coincides with a self-energy pole. Such large deviations of
FF-G$_0$W$_0$ from measured ionization potentials have been pointed out
before by Blase \emph{et al.}\cite{BlaseOlevano} for a number of
gas-phase molecules.

\begin{figure}
  \includegraphics[width=8.cm]{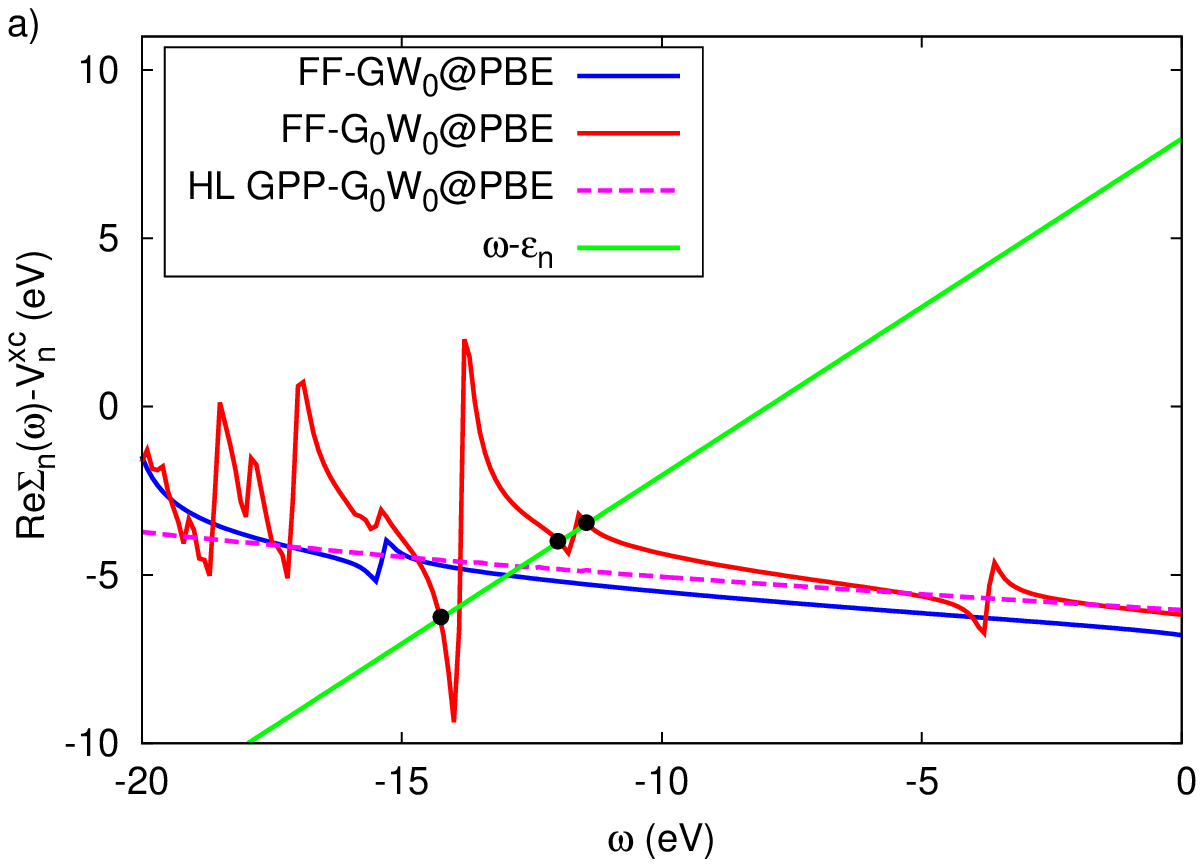} 
  \includegraphics[width=8.cm]{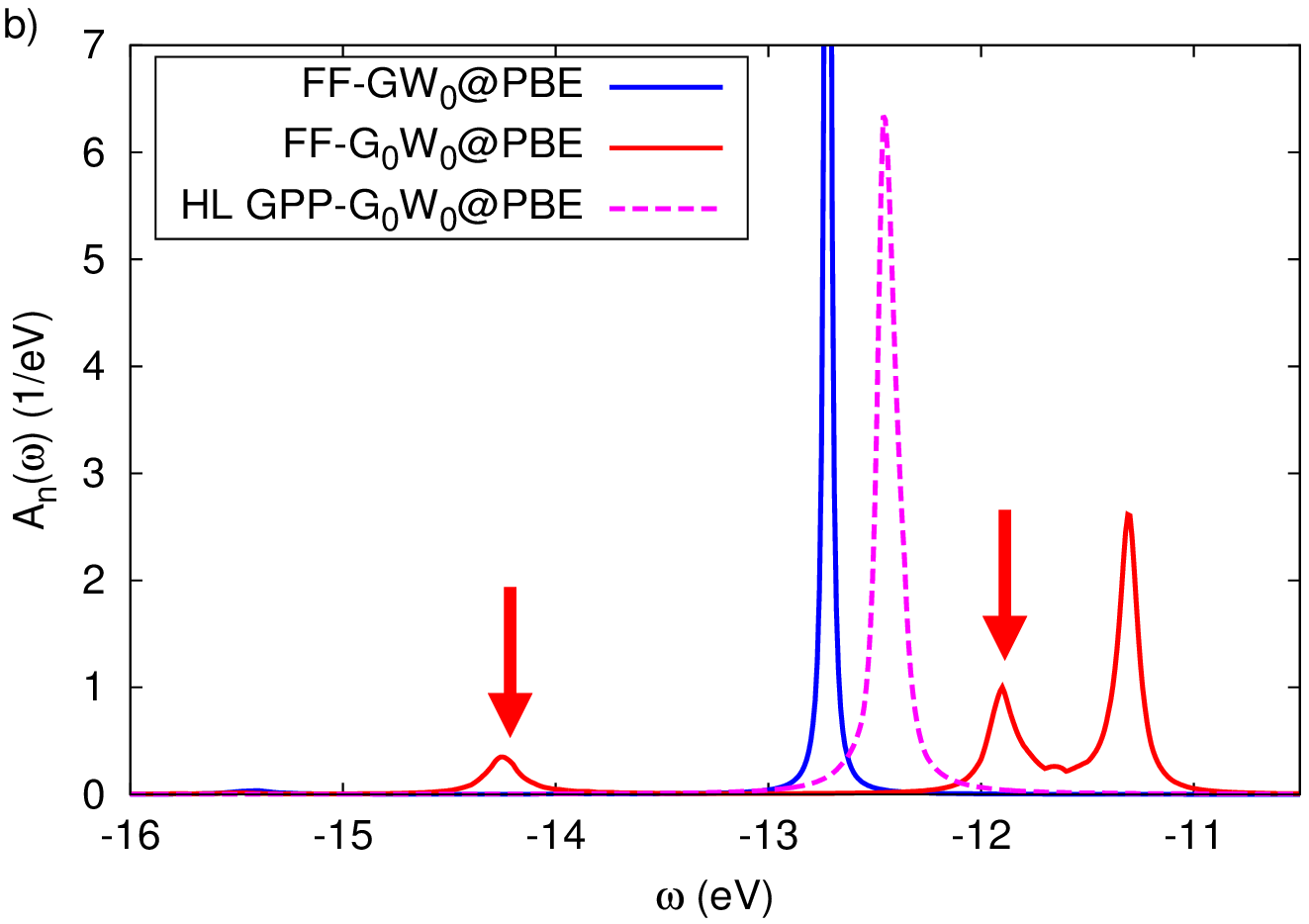}
  \caption{(a): Graphical solution of the quasiparticle equation for the HOMO of
    ozone. The quasiparticle energies $E_n$ are given by the values of
    $\omega$ at the intersections of $\omega-\epsilon_n$ and
    $\text{Re}\Sigma_n(\omega)-V^{xc}_n$. Shown are self energies from
    full-frequency $G_0W_0$ theory, full-frequency $GW_0$ theory and
    $G_0W_0$ theory with the generalized Hybertsen-Louie
    plasmon-pole approximation. All calculations employed a DFT-PBE
    starting point. (b): Resulting spectral functions for the HOMO of
    ozone. Arrows denote the position of shake-up features. Note that some
    solutions of the quasiparticle equation do not give rise to peaks
  in the spectral function because they are suppressed by strong peaks
in the imaginary part of the self energy. The solutions which give
rise to peaks in the spectral functions are marked by black dots.}
  \label{fig:O3}
\end{figure}

Finally, Fig.~\ref{fig:O3}(a) shows the self energy for the DFT-PBE
HOMO of ozone. Again, FF-$GW_0$ and HL GPP-$G_0W_0$ lead to excellent
agreement with experiment; however, FF-$G_0W_0$ yields a significant
discrepancy of $1.3$~eV because the quasiparticle energy is located in
the vicinity of a self-energy pole.

\begin{figure}
  \includegraphics[width=8.cm]{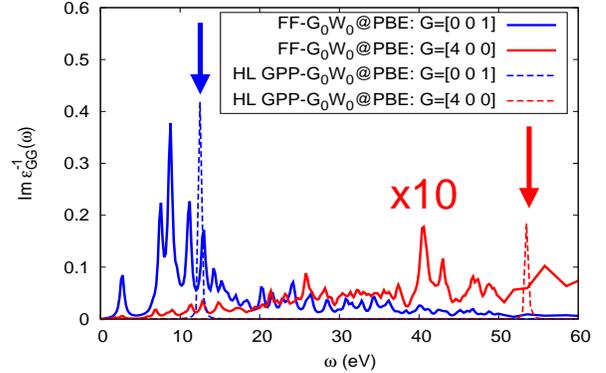}
  \caption{Imaginary part of the inverse dielectric matrix for the BeO
    molecule in a supercell calculations. Shown are the full-frequency
    result (FF) and the Hybertsen-Louie generalized plasmon-pole (HL
    GPP) model for
    $\bm{G}=\bm{G}'=[0 0 1]2\pi/a_0$ and $\bm{G}=\bm{G}'=[4 0
    0]2\pi/a_0$ (multiplied by a factor of 10) with $a_0$ denoting the
    linear dimension of the supercell. Arrows denote the positions of
    the effective excitations in the generalized plasmon-pole
    model. The molecular axis is along the z-direction.}
  \label{fig:BeO_epsinv}
\end{figure}

We thus find a strong correlation between the accuracy of the
self-energies poles and the accuracy of the resulting quasiparticle
energies. For all three molecules, FF-$GW_0$ and HL GPP-$G_0W_0$ lead
to self-energy poles separated by multiple electron volts from the
quasiparticle energy of the DFT-PBE HOMO and LUMO and give good
agreement with experiment. In contrast, we find significant
disagreement between experiment and FF-$G_0W_0$ results when the
quasiparticle energies are close to the incorrectly computed
self-energy poles. To understand the differences in the positions of
the self-energy poles, we express the FF-$G_0W_0$ self energy as the
sum of a bare exchange contribution and a frequency-dependent
correlation contribution given by
\begin{align}
  \langle m | \Sigma^c(\omega) | m\rangle = \sum_{nI}
  \frac{|V_{mnI}|^2}{\omega-\epsilon_n - \Omega_I
    \text{sgn}(\epsilon_n-\mu) + i\eta},
  \label{eq:fullfreq}
\end{align}
where $\mu$ denotes the chemical potential and $V_{jnI}$ is a
fluctuation potential \cite{Tiago,HedinReview}. Also, $\Omega_I$ is a
pole of the screened interaction $W$ and corresponds to a neutral
excitation energy of the system\cite{HedinBook}. For molecular
systems, the poles of the screened interaction within the random-phase
approximation are typically quite close to energy differences of the
DFT-PBE mean-field theory used to calculate $W$ \cite{Liu}.
Table~\ref{table:W_poles} shows that DFT-PBE energy differences agree
very well with experimental optical excitation energies in the three
molecules indicating that the screened interaction from DFT-PBE is
reasonable accurate. In contrast, Hartree-Fock energy differences
differ by multiple electron volts from experiment, as expected as
electron-hole attractions in optical excitations are neglected within
Hartree-Fock theory.

According to Eq.~\eqref{eq:fullfreq}, the FF-$G_0W_0$ self-energy
poles occur at $\omega=\epsilon_n-\Omega_I$ (if $n$ is an occupied
state). Even if the values of $\Omega_I$ were accurate, the
FF-$G_0W_0$ self-energy poles would be incorrectly positioned if the
mean-field energies $\epsilon_n$ differ from the quasiparticle
energies. In our approximate FF-GW$_0$ method, the DFT-PBE orbital
energies are replaced by Hartree-Fock energies, which are closer to
the correct quasiparticle energies and more negative by multiple
electron volts (see Table~\ref{table:energies}). This FF-$GW_0$
approach thus moves the self-energy poles to \emph{more negative}
energies. In the HL GPP-$G_0W_0$ method, DFT-PBE energies are used in
Eq.~\eqref{eq:fullfreq}, but for each $W_{\bm{G}\bm{G'}}$ all poles
are replaced by a single effective pole. To conserve sum
rules\cite{LouieHybertsen}, the energy of the effective pole must be
larger than the smallest $\Omega_I$, see
Figure~\ref{fig:BeO_epsinv}. In effect, this also results in a shift
of the self-energy poles to more negative energies. We thus find that
\emph{different reasons} are responsible for the shift of the
self-energy poles to more appropriate values in FF-GW$_0$ and HL
GPP-$G_0W_0$ approaches. We note that while the resulting self
energies agree quite well in the vicinity of the quasiparticle
solution, they disagree at higher energies where shake-up structures
are important. This could result in inaccuracies of the generalized
plasmon-pole approximations for the so-called inner valence
states\cite{cederbaum}.

\begin{table}
  \setlength{\doublerulesep}{1\doublerulesep}
  \setlength{\tabcolsep}{2\tabcolsep}
  \caption{Comparison of lowest experimental neutral singlet excitation energies of the
    molecules \cite{BeOexp,O3_grimme,CH4_miller,CH4_Moe,CH4_Sun} with energy differences from density-functional
    theory (DFT-PBE) and Hartree-Fock (HF) calculations. The neutral
    excitation energies are the poles of the screened interaction. All energies are given in eV.}
  \begin{ruledtabular}
    
    \begin{tabular}{c | c c c }
     &  DFT-PBE & HF & Exp. \\
      \hline
     CH$_4$ & 8.64 & 14.03 & 9.87-10.5\\
      BeO &  1.41 & 10.47 & 1.48\\
       O$_3$ & 1.80 & 13.24 & 2.0\\
    \end{tabular}

  \end{ruledtabular}
  \label{table:W_poles}
\end{table}

For unoccupied states in the sum in Eq.~\eqref{eq:fullfreq}, the
self-energy poles are located at $\omega=\epsilon_n+\Omega_I$. The
orbital energies in Hartree-Fock are again closer to the true
quasiparticle energies than those from DFT-PBE (see
Table~\ref{table:energies}), resulting in a shift of the self-energy
poles to more positive energies. The increase of the effective
$\Omega_I$ in the HL GPP-$G_0W_0$ theory has the same effect.  The
above discussion shows that use of FF-$G_0W_0$ is particularly
problematic for molecules with a small DFT-PBE HOMO-LUMO gap,
resulting in self-energy poles in the vicinity of the quasiparticle
energy.

Finally, we discuss the physical meaning of the singular structures in
the self energy. These poles give rise to additional peaks in the
spectral function [see Fig.~\ref{fig:O3}(b)] describing so-called
shake-up processes where an electron-hole pair is excited \emph{in
  addition} to a quasiparticle \cite{cederbaum}.  Also, in electronic
systems with open shells, the self-energy poles are responsible for
the \emph{multiplet structure} arising from the coupling of angular
momenta of the outer valence shell and of the hole left behind in the
photoemission process\cite{LischnerOpen}. In extended systems,
additional features in spectral functions arising from the shake-up of
plasmon modes, known as plasmon satellites, have received much
attention recently \cite{LischnerVigil,guzzo,Rotenberg2}.

\emph{Conclusions}.---We have computed self energies and quasiparticle
properties for three molecules using three approximate GW methods
employed a DFT-PBE mean-field starting point. Results of the
full-frequency $G_0W_0$ approximation can differ significantly (by
more than 1 eV) from experimental findings. We have traced this
failure of the full-frequency $G_0W_0$ method to the occurrence of
inaccurate self-energy poles in the vicinity of the quasiparticle
energy. Both a full-frequency $GW_0$ method and $G_0W_0$ with the
generalized plasmon-pole approximation shift the self-energy poles
away from the quasiparticle energies and lead to excellent agreement
with experiment. The generalized plasmon-pole model is therefore a
valuable approximation for molecular systems reducing the
computational cost significantly compared to full-frequency
self-consistent approaches. We expect that the effects of self
consistency are important for a wide range of molecules, particularly
those with a mean-field HOMO-LUMO gap of similar or smaller size than
the typical quasiparticle shifts.

\emph{Acknowledgments.---} This work was supported by NSF Grant
No. DMR10-1006184 (theoretical analysis) and by the SciDAC Program on
Excited State Phenomena (methods and software developments) and Theory
Program (GW calculations) funded by the U. S. Department of Energy,
the Office of Basic Energy Sciences and of Advanced Scientific
Computing Research, under Contract No. DE-AC02-05CH11231 at the
Lawrence Berkeley National Laboratory. S.G.L. acknowledges support by
a Simons Foundation Fellowship in Theoretical Physics. Computational
resources have been provided by the DOE at NERSC. Portions of this
work were carried out at the Molecular Foundry, also supported by the
US Department of Energy through the Office of Basic Energy Sciences.

\bibliography{paper}

\begin{thebibliography}{40}
\expandafter\ifx\csname natexlab\endcsname\relax\def\natexlab#1{#1}\fi
\expandafter\ifx\csname bibnamefont\endcsname\relax
  \def\bibnamefont#1{#1}\fi
\expandafter\ifx\csname bibfnamefont\endcsname\relax
  \def\bibfnamefont#1{#1}\fi
\expandafter\ifx\csname citenamefont\endcsname\relax
  \def\citenamefont#1{#1}\fi
\expandafter\ifx\csname url\endcsname\relax
  \def\url#1{\texttt{#1}}\fi
\expandafter\ifx\csname urlprefix\endcsname\relax\def\urlprefix{URL }\fi
\providecommand{\bibinfo}[2]{#2}
\providecommand{\eprint}[2][]{\url{#2}}

\bibitem[{\citenamefont{Lischner et~al.}(2013)\citenamefont{Lischner,
  Vigil-Fowler, and Louie}}]{LischnerVigil}
\bibinfo{author}{\bibfnamefont{J.}~\bibnamefont{Lischner}},
  \bibinfo{author}{\bibfnamefont{D.}~\bibnamefont{Vigil-Fowler}},
  \bibnamefont{and} \bibinfo{author}{\bibfnamefont{S.~G.} \bibnamefont{Louie}},
  \bibinfo{journal}{Phys. Rev. Lett.} \textbf{\bibinfo{volume}{110}},
  \bibinfo{pages}{146801} (\bibinfo{year}{2013}).

\bibitem[{\citenamefont{Gr\"{u}neis et~al.}(2008)\citenamefont{Gr\"{u}neis,
  Attaccalite, Pichler, Zabolotnyy, Shiozawa, Molodtsov, Inosov, Koitzsch,
  Knupfer, Schiessling et~al.}}]{Rubio}
\bibinfo{author}{\bibfnamefont{A.}~\bibnamefont{Gr\"{u}neis}},
  \bibinfo{author}{\bibfnamefont{C.}~\bibnamefont{Attaccalite}},
  \bibinfo{author}{\bibfnamefont{T.}~\bibnamefont{Pichler}},
  \bibinfo{author}{\bibfnamefont{V.}~\bibnamefont{Zabolotnyy}},
  \bibinfo{author}{\bibfnamefont{H.}~\bibnamefont{Shiozawa}},
  \bibinfo{author}{\bibfnamefont{S.~L.} \bibnamefont{Molodtsov}},
  \bibinfo{author}{\bibfnamefont{D.}~\bibnamefont{Inosov}},
  \bibinfo{author}{\bibfnamefont{A.}~\bibnamefont{Koitzsch}},
  \bibinfo{author}{\bibfnamefont{M.}~\bibnamefont{Knupfer}},
  \bibinfo{author}{\bibfnamefont{J.}~\bibnamefont{Schiessling}},
  \bibnamefont{et~al.}, \bibinfo{journal}{Phys. Rev. Lett.}
  \textbf{\bibinfo{volume}{100}}, \bibinfo{pages}{037601}
  (\bibinfo{year}{2008}).

\bibitem[{\citenamefont{Dial et~al.}(2012)\citenamefont{Dial, Ashoori,
  Pfeiffer, and West}}]{Dial}
\bibinfo{author}{\bibfnamefont{O.~E.} \bibnamefont{Dial}},
  \bibinfo{author}{\bibfnamefont{R.~C.} \bibnamefont{Ashoori}},
  \bibinfo{author}{\bibfnamefont{L.~N.} \bibnamefont{Pfeiffer}},
  \bibnamefont{and} \bibinfo{author}{\bibfnamefont{K.~W.} \bibnamefont{West}},
  \bibinfo{journal}{Phys. Rev. B} \textbf{\bibinfo{volume}{85}},
  \bibinfo{pages}{081306(R)} (\bibinfo{year}{2012}).

\bibitem[{\citenamefont{Cheng et~al.}(2011)\citenamefont{Cheng, Skouta,
  Vasquez, Widawsky, Schneebeli, Chen, Hybertsen, Breslow, and
  Venkataraman}}]{MolTransport}
\bibinfo{author}{\bibfnamefont{Z.-L.} \bibnamefont{Cheng}},
  \bibinfo{author}{\bibfnamefont{R.}~\bibnamefont{Skouta}},
  \bibinfo{author}{\bibfnamefont{H.}~\bibnamefont{Vasquez}},
  \bibinfo{author}{\bibfnamefont{J.~R.} \bibnamefont{Widawsky}},
  \bibinfo{author}{\bibfnamefont{S.}~\bibnamefont{Schneebeli}},
  \bibinfo{author}{\bibfnamefont{W.}~\bibnamefont{Chen}},
  \bibinfo{author}{\bibfnamefont{M.~S.} \bibnamefont{Hybertsen}},
  \bibinfo{author}{\bibfnamefont{R.}~\bibnamefont{Breslow}}, \bibnamefont{and}
  \bibinfo{author}{\bibfnamefont{L.}~\bibnamefont{Venkataraman}},
  \bibinfo{journal}{Nature Nanotechnology} \textbf{\bibinfo{volume}{6}},
  \bibinfo{pages}{353} (\bibinfo{year}{2011}).

\bibitem[{\citenamefont{Rohlfing and Louie}(1998)}]{LouieRohlfing}
\bibinfo{author}{\bibfnamefont{M.}~\bibnamefont{Rohlfing}} \bibnamefont{and}
  \bibinfo{author}{\bibfnamefont{S.~G.} \bibnamefont{Louie}},
  \bibinfo{journal}{Phys. Rev. Lett.} \textbf{\bibinfo{volume}{81}},
  \bibinfo{pages}{2312} (\bibinfo{year}{1998}).

\bibitem[{\citenamefont{Hedin and Lundqvist}(1969)}]{HedinBook}
\bibinfo{author}{\bibfnamefont{L.}~\bibnamefont{Hedin}} \bibnamefont{and}
  \bibinfo{author}{\bibfnamefont{S.}~\bibnamefont{Lundqvist}},
  \bibinfo{journal}{Solid State Physics} \textbf{\bibinfo{volume}{23}},
  \bibinfo{pages}{1} (\bibinfo{year}{1969}).

\bibitem[{\citenamefont{Hybertsen and Louie}(1986)}]{LouieHybertsen}
\bibinfo{author}{\bibfnamefont{M.~S.} \bibnamefont{Hybertsen}}
  \bibnamefont{and} \bibinfo{author}{\bibfnamefont{S.~G.} \bibnamefont{Louie}},
  \bibinfo{journal}{Phys. Rev. B} \textbf{\bibinfo{volume}{34}},
  \bibinfo{pages}{5390} (\bibinfo{year}{1986}).

\bibitem[{\citenamefont{Godby and Needs}(1989)}]{GodbyNeeds}
\bibinfo{author}{\bibfnamefont{R.~W.} \bibnamefont{Godby}} \bibnamefont{and}
  \bibinfo{author}{\bibfnamefont{R.~J.} \bibnamefont{Needs}},
  \bibinfo{journal}{Phys. Rev. Lett.} \textbf{\bibinfo{volume}{62}},
  \bibinfo{pages}{1169} (\bibinfo{year}{1989}).

\bibitem[{\citenamefont{von~der Linden and Horsch}(1988)}]{LindenHorsch}
\bibinfo{author}{\bibfnamefont{W.}~\bibnamefont{von~der Linden}}
  \bibnamefont{and} \bibinfo{author}{\bibfnamefont{P.}~\bibnamefont{Horsch}},
  \bibinfo{journal}{Phys. Rev. B} \textbf{\bibinfo{volume}{37}},
  \bibinfo{pages}{8351} (\bibinfo{year}{1988}).

\bibitem[{\citenamefont{Hahn et~al.}(2005)\citenamefont{Hahn, Schmidt, and
  Bechstedt}}]{Bechstedt}
\bibinfo{author}{\bibfnamefont{P.~H.} \bibnamefont{Hahn}},
  \bibinfo{author}{\bibfnamefont{W.~G.} \bibnamefont{Schmidt}},
  \bibnamefont{and}
  \bibinfo{author}{\bibfnamefont{F.}~\bibnamefont{Bechstedt}},
  \bibinfo{journal}{Phys. Rev. B} \textbf{\bibinfo{volume}{72}},
  \bibinfo{pages}{245425} (\bibinfo{year}{2005}).

\bibitem[{\citenamefont{Lischner et~al.}(2012)\citenamefont{Lischner, Deslippe,
  Jain, and Louie}}]{LischnerOpen}
\bibinfo{author}{\bibfnamefont{J.}~\bibnamefont{Lischner}},
  \bibinfo{author}{\bibfnamefont{J.}~\bibnamefont{Deslippe}},
  \bibinfo{author}{\bibfnamefont{M.}~\bibnamefont{Jain}}, \bibnamefont{and}
  \bibinfo{author}{\bibfnamefont{S.~G.} \bibnamefont{Louie}},
  \bibinfo{journal}{Phys. Rev. Lett.} \textbf{\bibinfo{volume}{109}},
  \bibinfo{pages}{036406} (\bibinfo{year}{2012}).

\bibitem[{\citenamefont{Sharifzadeh et~al.}(2012)\citenamefont{Sharifzadeh,
  Tamblyn, Doak, Darancet, and Neaton}}]{SaharNeaton}
\bibinfo{author}{\bibfnamefont{S.}~\bibnamefont{Sharifzadeh}},
  \bibinfo{author}{\bibfnamefont{I.}~\bibnamefont{Tamblyn}},
  \bibinfo{author}{\bibfnamefont{P.}~\bibnamefont{Doak}},
  \bibinfo{author}{\bibfnamefont{P.~T.} \bibnamefont{Darancet}},
  \bibnamefont{and} \bibinfo{author}{\bibfnamefont{J.~B.}
  \bibnamefont{Neaton}}, \bibinfo{journal}{Eur. Phys. J. B}
  \textbf{\bibinfo{volume}{85}}, \bibinfo{pages}{323} (\bibinfo{year}{2012}).

\bibitem[{\citenamefont{Grossman et~al.}(2001)\citenamefont{Grossman, Rohlfing,
  Mitas, Louie, and Cohen}}]{SteveGrossmann}
\bibinfo{author}{\bibfnamefont{J.~C.} \bibnamefont{Grossman}},
  \bibinfo{author}{\bibfnamefont{M.}~\bibnamefont{Rohlfing}},
  \bibinfo{author}{\bibfnamefont{L.}~\bibnamefont{Mitas}},
  \bibinfo{author}{\bibfnamefont{S.~G.} \bibnamefont{Louie}}, \bibnamefont{and}
  \bibinfo{author}{\bibfnamefont{M.~L.} \bibnamefont{Cohen}},
  \bibinfo{journal}{Phys. Rev. Lett.} \textbf{\bibinfo{volume}{86}},
  \bibinfo{pages}{472} (\bibinfo{year}{2001}).

\bibitem[{\citenamefont{Neaton et~al.}(2006)\citenamefont{Neaton, Hybertsen,
  and Louie}}]{NeatonLouie}
\bibinfo{author}{\bibfnamefont{J.~B.} \bibnamefont{Neaton}},
  \bibinfo{author}{\bibfnamefont{M.~S.} \bibnamefont{Hybertsen}},
  \bibnamefont{and} \bibinfo{author}{\bibfnamefont{S.~G.} \bibnamefont{Louie}},
  \bibinfo{journal}{Phys. Rev. Lett.} \textbf{\bibinfo{volume}{97}},
  \bibinfo{pages}{216405} (\bibinfo{year}{2006}).

\bibitem[{\citenamefont{Blase et~al.}(2011)\citenamefont{Blase, Attaccalite,
  and Olevano}}]{BlaseOlevano}
\bibinfo{author}{\bibfnamefont{X.}~\bibnamefont{Blase}},
  \bibinfo{author}{\bibfnamefont{C.}~\bibnamefont{Attaccalite}},
  \bibnamefont{and} \bibinfo{author}{\bibfnamefont{V.}~\bibnamefont{Olevano}},
  \bibinfo{journal}{Phys. Rev. B} \textbf{\bibinfo{volume}{83}},
  \bibinfo{pages}{115103} (\bibinfo{year}{2011}).

\bibitem[{\citenamefont{Foerster et~al.}(2011)\citenamefont{Foerster, Koval,
  and Sanchez-Portal}}]{SanchezPortal}
\bibinfo{author}{\bibfnamefont{D.}~\bibnamefont{Foerster}},
  \bibinfo{author}{\bibfnamefont{P.}~\bibnamefont{Koval}}, \bibnamefont{and}
  \bibinfo{author}{\bibfnamefont{D.}~\bibnamefont{Sanchez-Portal}},
  \bibinfo{journal}{J. Chem. Phys.} \textbf{\bibinfo{volume}{135}},
  \bibinfo{pages}{074105} (\bibinfo{year}{2011}).

\bibitem[{\citenamefont{Thygesen and Rubio}(2009)}]{ThygesenRubio}
\bibinfo{author}{\bibfnamefont{K.~S.} \bibnamefont{Thygesen}} \bibnamefont{and}
  \bibinfo{author}{\bibfnamefont{A.}~\bibnamefont{Rubio}},
  \bibinfo{journal}{Phys. Rev. Lett.} \textbf{\bibinfo{volume}{102}},
  \bibinfo{pages}{046802} (\bibinfo{year}{2009}).

\bibitem[{\citenamefont{H{\"u}ser et~al.}(2013)\citenamefont{H{\"u}ser, Olsen,
  and Thygesen}}]{ThygesenOlsen}
\bibinfo{author}{\bibfnamefont{F.}~\bibnamefont{H{\"u}ser}},
  \bibinfo{author}{\bibfnamefont{T.}~\bibnamefont{Olsen}}, \bibnamefont{and}
  \bibinfo{author}{\bibfnamefont{K.~S.} \bibnamefont{Thygesen}},
  \bibinfo{journal}{Phys. Rev. B} \textbf{\bibinfo{volume}{87}},
  \bibinfo{pages}{235132} (\bibinfo{year}{2013}).

\bibitem[{\citenamefont{K\"{o}rzd\"{o}rfer and Marom}(2012)}]{Marom}
\bibinfo{author}{\bibfnamefont{T.}~\bibnamefont{K\"{o}rzd\"{o}rfer}}
  \bibnamefont{and} \bibinfo{author}{\bibfnamefont{N.}~\bibnamefont{Marom}},
  \bibinfo{journal}{Phys. Rev. B} \textbf{\bibinfo{volume}{86}},
  \bibinfo{pages}{041110(R)} (\bibinfo{year}{2012}).

\bibitem[{\citenamefont{Rostgaard et~al.}(2010)\citenamefont{Rostgaard,
  Jacobson, and Thygesen}}]{ThygesenRostgaard}
\bibinfo{author}{\bibfnamefont{C.}~\bibnamefont{Rostgaard}},
  \bibinfo{author}{\bibfnamefont{K.~W.} \bibnamefont{Jacobson}},
  \bibnamefont{and} \bibinfo{author}{\bibfnamefont{K.~S.}
  \bibnamefont{Thygesen}}, \bibinfo{journal}{Phys. Rev. B}
  \textbf{\bibinfo{volume}{81}}, \bibinfo{pages}{085103}
  (\bibinfo{year}{2010}).

\bibitem[{\citenamefont{Shirley and Martin}(1993)}]{ShirleyMartin}
\bibinfo{author}{\bibfnamefont{E.~L.} \bibnamefont{Shirley}} \bibnamefont{and}
  \bibinfo{author}{\bibfnamefont{R.~M.} \bibnamefont{Martin}},
  \bibinfo{journal}{Phys. Rev. B} \textbf{\bibinfo{volume}{47}},
  \bibinfo{pages}{15404} (\bibinfo{year}{1993}).

\bibitem[{\citenamefont{Caruso et~al.}(2012)\citenamefont{Caruso, Rinke, Ren,
  Scheffler, and Rubio}}]{Caruso}
\bibinfo{author}{\bibfnamefont{F.}~\bibnamefont{Caruso}},
  \bibinfo{author}{\bibfnamefont{P.}~\bibnamefont{Rinke}},
  \bibinfo{author}{\bibfnamefont{X.}~\bibnamefont{Ren}},
  \bibinfo{author}{\bibfnamefont{M.}~\bibnamefont{Scheffler}},
  \bibnamefont{and} \bibinfo{author}{\bibfnamefont{A.}~\bibnamefont{Rubio}},
  \bibinfo{journal}{Phys. Rev. B} \textbf{\bibinfo{volume}{86}},
  \bibinfo{pages}{081102(R)} (\bibinfo{year}{2012}).

\bibitem[{\citenamefont{Stan et~al.}(2009)\citenamefont{Stan, Dahlen, and van
  Leeuwen}}]{vanLeeuwen}
\bibinfo{author}{\bibfnamefont{A.}~\bibnamefont{Stan}},
  \bibinfo{author}{\bibfnamefont{N.~E.} \bibnamefont{Dahlen}},
  \bibnamefont{and} \bibinfo{author}{\bibfnamefont{R.}~\bibnamefont{van
  Leeuwen}}, \bibinfo{journal}{J. Chem. Phys.} \textbf{\bibinfo{volume}{130}},
  \bibinfo{pages}{114105} (\bibinfo{year}{2009}).

\bibitem[{\citenamefont{Cederbaum et~al.}(1980)\citenamefont{Cederbaum, Domcke,
  Schirmer, and von Niessen}}]{cederbaum}
\bibinfo{author}{\bibfnamefont{L.~S.} \bibnamefont{Cederbaum}},
  \bibinfo{author}{\bibfnamefont{W.}~\bibnamefont{Domcke}},
  \bibinfo{author}{\bibfnamefont{J.}~\bibnamefont{Schirmer}}, \bibnamefont{and}
  \bibinfo{author}{\bibfnamefont{W.}~\bibnamefont{von Niessen}},
  \bibinfo{journal}{Physica Scripta} \textbf{\bibinfo{volume}{21}},
  \bibinfo{pages}{481} (\bibinfo{year}{1980}).

\bibitem[{\citenamefont{Holm and von Barth}(2004)}]{Holm}
\bibinfo{author}{\bibfnamefont{B.}~\bibnamefont{Holm}} \bibnamefont{and}
  \bibinfo{author}{\bibfnamefont{U.}~\bibnamefont{von Barth}},
  \bibinfo{journal}{Physica Scripta} \textbf{\bibinfo{volume}{T109}},
  \bibinfo{pages}{135} (\bibinfo{year}{2004}).

\bibitem[{\citenamefont{Giannozzi et~al.}(2009)\citenamefont{Giannozzi, Baroni,
  Bonini, Calandra, Car, Cavazzoni, Ceresoli, Chiarotti, Cococcioni, Dabo
  et~al.}}]{QuantumEspresso}
\bibinfo{author}{\bibfnamefont{P.}~\bibnamefont{Giannozzi}},
  \bibinfo{author}{\bibfnamefont{S.}~\bibnamefont{Baroni}},
  \bibinfo{author}{\bibfnamefont{N.}~\bibnamefont{Bonini}},
  \bibinfo{author}{\bibfnamefont{M.}~\bibnamefont{Calandra}},
  \bibinfo{author}{\bibfnamefont{R.}~\bibnamefont{Car}},
  \bibinfo{author}{\bibfnamefont{C.}~\bibnamefont{Cavazzoni}},
  \bibinfo{author}{\bibfnamefont{D.}~\bibnamefont{Ceresoli}},
  \bibinfo{author}{\bibfnamefont{G.~L.} \bibnamefont{Chiarotti}},
  \bibinfo{author}{\bibfnamefont{M.}~\bibnamefont{Cococcioni}},
  \bibinfo{author}{\bibfnamefont{I.}~\bibnamefont{Dabo}}, \bibnamefont{et~al.},
  \bibinfo{journal}{Journal of Physics: Condensed Matter}
  \textbf{\bibinfo{volume}{21}}, \bibinfo{pages}{395502}
  (\bibinfo{year}{2009}).

\bibitem[{\citenamefont{Tiago and Chelikowsky}(2006)}]{Tiago}
\bibinfo{author}{\bibfnamefont{M.~L.} \bibnamefont{Tiago}} \bibnamefont{and}
  \bibinfo{author}{\bibfnamefont{J.~R.} \bibnamefont{Chelikowsky}},
  \bibinfo{journal}{Phys. Rev. B} \textbf{\bibinfo{volume}{73}},
  \bibinfo{pages}{205334} (\bibinfo{year}{2006}).

\bibitem[{\citenamefont{Deslippe et~al.}(2012)\citenamefont{Deslippe,
  Samsonidze, Strubbe, Jain, Cohen, and Louie}}]{BGWpaper}
\bibinfo{author}{\bibfnamefont{J.}~\bibnamefont{Deslippe}},
  \bibinfo{author}{\bibfnamefont{G.}~\bibnamefont{Samsonidze}},
  \bibinfo{author}{\bibfnamefont{D.~A.} \bibnamefont{Strubbe}},
  \bibinfo{author}{\bibfnamefont{M.}~\bibnamefont{Jain}},
  \bibinfo{author}{\bibfnamefont{M.~L.} \bibnamefont{Cohen}}, \bibnamefont{and}
  \bibinfo{author}{\bibfnamefont{S.~G.} \bibnamefont{Louie}},
  \bibinfo{journal}{Comput. Phys. Commun.} \textbf{\bibinfo{volume}{183}},
  \bibinfo{pages}{1269} (\bibinfo{year}{2012}).

\bibitem[{\citenamefont{Deslippe et~al.}(2013)\citenamefont{Deslippe,
  Samsonidze, Jain, Cohen, and Louie}}]{StaticRemainder}
\bibinfo{author}{\bibfnamefont{J.}~\bibnamefont{Deslippe}},
  \bibinfo{author}{\bibfnamefont{G.}~\bibnamefont{Samsonidze}},
  \bibinfo{author}{\bibfnamefont{M.}~\bibnamefont{Jain}},
  \bibinfo{author}{\bibfnamefont{M.~L.} \bibnamefont{Cohen}}, \bibnamefont{and}
  \bibinfo{author}{\bibfnamefont{S.~G.} \bibnamefont{Louie}},
  \bibinfo{journal}{Phys. Rev. B} \textbf{\bibinfo{volume}{87}},
  \bibinfo{pages}{165124} (\bibinfo{year}{2013}).

\bibitem[{\citenamefont{Ismail-Beigi}(2006)}]{sohrab}
\bibinfo{author}{\bibfnamefont{S.}~\bibnamefont{Ismail-Beigi}},
  \bibinfo{journal}{Phys. Rev. B} \textbf{\bibinfo{volume}{73}},
  \bibinfo{pages}{233103} (\bibinfo{year}{2006}).

\bibitem[{\citenamefont{Weast and Astle}()}]{CRC}
\bibinfo{author}{\bibfnamefont{R.~C.} \bibnamefont{Weast}} \bibnamefont{and}
  \bibinfo{author}{\bibfnamefont{M.~J.} \bibnamefont{Astle}},
  \bibinfo{journal}{CRC Handbook of Chemistry and Physics, 92nd ed.}  (????).

\bibitem[{\citenamefont{Hedin}(1999)}]{HedinReview}
\bibinfo{author}{\bibfnamefont{L.}~\bibnamefont{Hedin}}, \bibinfo{journal}{J.
  Phys.: Condens. Matter} \textbf{\bibinfo{volume}{11}}, \bibinfo{pages}{489}
  (\bibinfo{year}{1999}).

\bibitem[{\citenamefont{Liu et~al.}(2014)\citenamefont{Liu, Lin, Vigil-Fowler,
  Lischner, Kemper, Sharifzadeh, da~Jornada, Deslipee, Yang, Neaton
  et~al.}}]{Liu}
\bibinfo{author}{\bibfnamefont{F.}~\bibnamefont{Liu}},
  \bibinfo{author}{\bibfnamefont{L.}~\bibnamefont{Lin}},
  \bibinfo{author}{\bibfnamefont{D.}~\bibnamefont{Vigil-Fowler}},
  \bibinfo{author}{\bibfnamefont{J.}~\bibnamefont{Lischner}},
  \bibinfo{author}{\bibfnamefont{A.~F.} \bibnamefont{Kemper}},
  \bibinfo{author}{\bibfnamefont{S.}~\bibnamefont{Sharifzadeh}},
  \bibinfo{author}{\bibfnamefont{F.~H.} \bibnamefont{da~Jornada}},
  \bibinfo{author}{\bibfnamefont{J.}~\bibnamefont{Deslipee}},
  \bibinfo{author}{\bibfnamefont{C.}~\bibnamefont{Yang}},
  \bibinfo{author}{\bibfnamefont{J.~B.} \bibnamefont{Neaton}},
  \bibnamefont{et~al.}, \bibinfo{journal}{arXiv:1402.5433}
  (\bibinfo{year}{2014}).

\bibitem[{\citenamefont{Verhaegen and Richards}(1966)}]{BeOexp}
\bibinfo{author}{\bibfnamefont{G.}~\bibnamefont{Verhaegen}} \bibnamefont{and}
  \bibinfo{author}{\bibfnamefont{W.~G.} \bibnamefont{Richards}},
  \bibinfo{journal}{J. Chem. Phys.} \textbf{\bibinfo{volume}{45}},
  \bibinfo{pages}{1828} (\bibinfo{year}{1966}).

\bibitem[{\citenamefont{Grimme and Waletzke}(1999)}]{O3_grimme}
\bibinfo{author}{\bibfnamefont{S.}~\bibnamefont{Grimme}} \bibnamefont{and}
  \bibinfo{author}{\bibfnamefont{M.}~\bibnamefont{Waletzke}},
  \bibinfo{journal}{J. Chem. Phys.} \textbf{\bibinfo{volume}{111}},
  \bibinfo{pages}{5645} (\bibinfo{year}{1999}).

\bibitem[{\citenamefont{Bowman and Miller}(1965)}]{CH4_miller}
\bibinfo{author}{\bibfnamefont{C.~R.} \bibnamefont{Bowman}} \bibnamefont{and}
  \bibinfo{author}{\bibfnamefont{W.~D.} \bibnamefont{Miller}},
  \bibinfo{journal}{J. Chem. Phys.} \textbf{\bibinfo{volume}{42}},
  \bibinfo{pages}{681} (\bibinfo{year}{1965}).

\bibitem[{\citenamefont{Moe and Duncan}(1952)}]{CH4_Moe}
\bibinfo{author}{\bibfnamefont{G.}~\bibnamefont{Moe}} \bibnamefont{and}
  \bibinfo{author}{\bibfnamefont{A.~B.~F.} \bibnamefont{Duncan}},
  \bibinfo{journal}{J. Am. Chem.} \textbf{\bibinfo{volume}{74}},
  \bibinfo{pages}{3140} (\bibinfo{year}{1952}).

\bibitem[{\citenamefont{Sun and Weissler}(1955)}]{CH4_Sun}
\bibinfo{author}{\bibfnamefont{H.}~\bibnamefont{Sun}} \bibnamefont{and}
  \bibinfo{author}{\bibfnamefont{G.~L.} \bibnamefont{Weissler}},
  \bibinfo{journal}{J. Chem. Phys.} \textbf{\bibinfo{volume}{23}},
  \bibinfo{pages}{1160} (\bibinfo{year}{1955}).

\bibitem[{\citenamefont{Guzzo et~al.}(2011)\citenamefont{Guzzo, Lani, Sottile,
  Romaniello, Gatti, Kas, Rehr, Silly, Sirotti, and Reining}}]{guzzo}
\bibinfo{author}{\bibfnamefont{M.}~\bibnamefont{Guzzo}},
  \bibinfo{author}{\bibfnamefont{G.}~\bibnamefont{Lani}},
  \bibinfo{author}{\bibfnamefont{F.}~\bibnamefont{Sottile}},
  \bibinfo{author}{\bibfnamefont{P.}~\bibnamefont{Romaniello}},
  \bibinfo{author}{\bibfnamefont{M.}~\bibnamefont{Gatti}},
  \bibinfo{author}{\bibfnamefont{J.~J.} \bibnamefont{Kas}},
  \bibinfo{author}{\bibfnamefont{J.~J.} \bibnamefont{Rehr}},
  \bibinfo{author}{\bibfnamefont{M.~G.} \bibnamefont{Silly}},
  \bibinfo{author}{\bibfnamefont{F.}~\bibnamefont{Sirotti}}, \bibnamefont{and}
  \bibinfo{author}{\bibfnamefont{L.}~\bibnamefont{Reining}},
  \bibinfo{journal}{Phys. Rev. Lett.} \textbf{\bibinfo{volume}{107}},
  \bibinfo{pages}{166401} (\bibinfo{year}{2011}).

\bibitem[{\citenamefont{Bostwick et~al.}(2010)\citenamefont{Bostwick, Speck,
  Seyller, Horn, Polini, Asgari, MacDonald, and Rotenberg}}]{Rotenberg2}
\bibinfo{author}{\bibfnamefont{A.}~\bibnamefont{Bostwick}},
  \bibinfo{author}{\bibfnamefont{F.}~\bibnamefont{Speck}},
  \bibinfo{author}{\bibfnamefont{T.}~\bibnamefont{Seyller}},
  \bibinfo{author}{\bibfnamefont{K.}~\bibnamefont{Horn}},
  \bibinfo{author}{\bibfnamefont{M.}~\bibnamefont{Polini}},
  \bibinfo{author}{\bibfnamefont{R.}~\bibnamefont{Asgari}},
  \bibinfo{author}{\bibfnamefont{A.~H.} \bibnamefont{MacDonald}},
  \bibnamefont{and}
  \bibinfo{author}{\bibfnamefont{E.}~\bibnamefont{Rotenberg}},
  \bibinfo{journal}{Science} \textbf{\bibinfo{volume}{328}},
  \bibinfo{pages}{999} (\bibinfo{year}{2010}).

\end{thebibliography}
\end{document}